# QUANTIZED FAILURE CRITERIA AND INDIRECT OBSERVATION FOR PREDICTING THE NANOSCALE STRENGTH OF MATERIALS: THE EXAMPLE OF THE ULTRA NANO CRYSTALLINE DIAMOND

Nicola M. Pugno
Department of Structural Engineering, Politecnico di Torino,
Corso Duca degli Abruzzi 24, 10129, Torino, Italy.
nicola.pugno@polito.it

**Abstract**

In this paper theoretical and statistical/experimental criteria for determining the nanoscale strength of materials are proposed. In particular, quantized criteria in fracture mechanics, dynamic fracture mechanics and fatigue, as well as an experimental indirect observation of the nanoscale strength, are proposed. The increasing of the dynamic resistance and the role of a fractal crack surface formation are also rationalized. The analysis shows that materials can be sensitive to flaws also at nanoscale (as demonstrated for carbon nanotubes), in contrast to the conclusion of a recently published paper, and that the surfaces are weaker than the inner parts of a solid by a factor of ~10%. In addition, the proposed statistical/experimental procedure is applied for predicting the nanoscale strength of the ultrananocrystalline diamond (UNCD), an innovative material only recently developed.

Regarding UNCD, even if its strength has been successfully measured at the microscale as ~4GPa, its nanoscale strength is a parameter still missing in literature. In spite of this its prediction is fundamental, e.g., for developing future innovative micro- and nano-electromechanical systems (MEMS and NEMS) based on UNCD nanowires. After a short description of the micro-experiments available in literature, the observed size-effects on material strength are rationalized with the help of the fractal statistics, demonstrated to be equivalent to the Weibull statistics. The experimental observations fitted with the fractal statistics show coefficient of correlations basically identical to 1. From the data analysis, the first "experimental" estimation of the UNCD nanoscale (tensile) strength is deduced as ~23GPa for a hypothetical nanowire with specified geometry ($10 \times 10 \times 100 \text{nm}^3$), whereas the nanoscale bending strength is estimated as ~30GPa. A fractal exponent of ~2.5 and a Weibull modulus of ~10 are also derived. Such results are shown to be compatible with a theoretical estimation of the UNCD ideal strength, deduced as ~52GPa by applying Quantized Fracture Mechanics, assuming the fracture quantum coincident with the grain size of the UNCD (~3nm). In contrast to what has been observed at microscale, the doping of the UNCD, imposed to increase its electrical conductivity, as required for NEMS & MEMS applications, seems to increase its strength at nanoscale. Even if such a result has to be considered with caution, it would be a consequence of the variation of the dimension of the fractal domain in which the energy dissipation occurs, imposed by the doping.

# 1. Introduction

UNCD material (Gruen, 1999) has been recently developed at Argonne National Laboratory and possesses unique properties particularly suitable to the design of novel MEMS. The UNCD films are grown by microwave plasma enhanced chemical vapor deposition (MPCVD) synthesis method, involving a rich $CH_4$/Ar plasma chemistry, where $C_2$ dimers are the growth species derived from collision induced fragmentation of $CH_4$ molecules in Ar plasma. The UNCD film growth proceeds via the reactions $2CH_4 \rightarrow C_2H_2 + 3H_2$ & $C_2H_2 \rightarrow C_2 + H_2$ in atmosphere containing very small quantities of hydrogen. UNCD films can be doped with nitrogen using a $CH_4$ (1%Ar) gas mixture and nitrogen gas added. The doping increases the electrical conductivity of the material, as required in MEMS applications. Strength of undoped (Espinosa et al., 2003; Pugno, Peng and Espinosa, 2004) and doped (Peng, Espinosa and Moldovan, 2004) UNCD has been recently investigated at microscale considering 0, 5, 10, 15, and 20% of nitrogen gas added in the controlled atmosphere. We demonstrate that the observed size-effects can be successfully rationalized on the light of the fractal statistics (Carpinteri and Pugno, 2002, 2004a); it shows that an increasing in the nitrogen content corresponds to a lower dimension of the fractal domain in which the energy dissipation during stretching occurs, resulting at microscale in a more brittle behavior. From the experimental results, the measured strength of UNCD at microscale was found to be ~4GPa, strongly reduced (by a factor of about 50%) from the presence of the doping, with fracture toughness of ~4MPam$^{0.5}$ and Young's modulus of ~1TPa (Espinosa et al., 2003; Pugno, Peng and Espinosa 2004; Peng, Espinosa and Moldovan, 2004). Unfortunately, until now no information was deduced regarding the fundamental issue (especially for NEMS applications) of the UNCD nanoscale strength. In contrast to the limits of the experimental analysis, unable until now to investigate the UNCD at such a scale, we obtain here an indirect "experimental" estimation of the UNCD nanoscale (tensile) strength by virtue of fractal statistics (Carpinteri and Pugno 2002, 2004a). It is found to be of ~23GPa (the nanoscale bending strength is estimated as ~30GPa) for a hypothetical nanowire of given width $W$=10nm, thickness $t$=10nm and length $L$=100nm. A surprising additional result is that, from the statistical data analysis, it seems that the presence of the doping basically would not affect the nanoscale strength. Even if such a result must be taken with caution, it clearly shows that doped UNCD nanowires, can be considered good candidates for NEMS applications (Pugno 2004a,b).

In addition to this general statistical/experimental procedure for the indirect observation of the nanoscale strength, useful quantized criteria proposed in literature and new ones, here presented to complete the scenario, are discussed to treat static, dynamic, stable and unstable crack propagations, also at nanoscale. In particular, by applying Quantized Fracture Mechanics (QFM; Pugno and Ruoff, 2004a) we show that the UNCD nanoscale strength is compatible with its ideal strength, estimated as ~52GPa.

Finally, we demonstrate that nanotubes are strongly sensitive to flaws, in contrast to the conclusion reported in the title of the paper (however very interesting) by Gao et al. (2003) that "materials become insensitive to flaws at nanoscale". The "doubling" of the dynamic resistance, the role of a fractal crack surface formation and the prediction that surfaces are weaker than the inner parts of a solid are also deduced.

## 2. Strength of solids: quantized criteria based on energy, stress and strain for fracture mechanics, dynamics fracture mechanics and fatigue crack growth

According to continuum based fracture mechanics (Griffith, 1920), the strength of a structure can be computed by setting the stress-intensity factor $K$ (for specified geometry and applied load) equal to its critical value (the fracture toughness of the material $K_C$), i.e., $K = K_C$ (for crack propagation modes I, II or III). On the other hand, if the crack advancement is assumed to be quantized (Pugno, 2002), the criterion becomes (*Quantized Fracture Mechanics* (Pugno and Ruoff, 2004a)):

$$K^* = \sqrt{\left\langle K^2 \right\rangle_l^{l+\Delta l}} = K_C; \quad \text{Modes I,II,III} \tag{1}$$

where $K^*$ is the square root of the mean value of the square of the stress-intensity factor along a fracture quantum $\Delta l$, for a crack of length $l$. The effectiveness of this approach has been demonstrated at nanoscale (Pugno and Ruoff, 2004a) but also by fitting experimental results at larger size scales (Pugno and Ruoff, 2004a; Taylor, Cornetti and Pugno, 2004) where the theory can be also called Finite Fracture Mechanics. An application is given in Section 3.

Analogously, for dynamic loads the mean value must be considered also along the time quantum $\Delta t$, connected to the time $\Delta l/v$ -with $v$ crack speed- to generate the fracture quantum, i.e. (*Dynamic Quantized Fracture Mechanics* (Pugno and Ruoff, 2004b)):

$$K_d^* = \sqrt{\left\langle \left\langle K^2 \right\rangle_l^{l+\Delta l} \right\rangle_{t-\Delta t}^{t}} = K_C; \quad \text{Modes I,II,III} \tag{2}$$

Note that classical dynamic fracture mechanics would imply $K = K_{dC}$ with $K_{dC}$ an *a priory* unknown dynamic fracture initiation toughness, different from $K_C$, especially for severe loading rates, e.g., impacts; on the other hand, eq. (2) reproduces very well the experimental observations also for severe loading rates, see Pugno and Ruoff (2004b). An application is given in Section 4. For taking into account also negative stress-intensity factor regimes (crack closure, in addition to the crack opening assumed by Griffith) $K^2$ could be considered with the algebraic sign of $K$ (fracture does not occur if negative and positive stages compensate each other during the fracture and time quanta). In fact, according to Griffith $K^2 = K_C^2$ whereas in general, also for crack closure, $K = K_C$ (where $K$ can be positive or negative).

In contrast to classical fracture mechanics (limit case for $\Delta l = 0$), that can be applied only to "large" (crack length larger than the fracture quantum, if viceversa the crack is here defined as "short") and sharp (vanishing tip radius) cracks, quantized fracture mechanics has no restriction in treating defects with any size and shape (Pugno and Ruoff, 2004a). Furthermore, dynamic quantized fracture mechanics can treat also severe loading rates (e.g., impacts) in contrast to classical dynamic fracture mechanics (limit case for $\Delta t = 0$) that becomes not predictive for such cases, requiring an *ad hoc* dynamic fracture initiation toughness; it is identical to its static value (as must be) only the dynamic quantized fracture mechanics treatment.

Instead of a classical maximum stress ($\sigma_{max}$) criterion, i.e., $\sigma_{max} = \sigma_C$, where $\sigma_C$ is the strength of the material, the stress analog of the energy based criterion (1) must be written as (*Quantized Maximum Stress Criterion* (Neuber, 1958; Novozhilov 1969)):

$$\sigma^* = \langle \sigma_{tip} \rangle_0^{\Delta l} = \sigma_C \quad \text{Mode I; for Modes II, III: } \sigma \rightarrow \tau \qquad (3)$$

where $\sigma_{tip}$ is the (opening, for crack propagation mode I) normal stress field at the tip of a defect, where is located the origin of the reference system; for mode II or III the normal stress and strength are replaced by the corresponding shear stress $\tau$ and strength $\tau_C$. This criterion, the first "quantized" one, was introduced by Neuber (1958) (especially in fatigue) and by Novozhilov (1969) (for brittle fracture); in particular Novozhilov (that introduced the term "fracture quantum") assumed that the fracture quantum be coincident with the interatomic spacing; however, his school and his apprentices widely applied this method during 1970-1980 removing such hypothesis, thus assuming that the fracture quantum is not restricted to be the atomic spacing (Morozov, 1984). Only later other authors applied this modified criterion (Sewerin, 1998; Taylor, 1999; Carpinteri and Pugno 2004b).

For dynamic loads this criterion has to be rewritten as (that we could call *Dynamic Quantized Maximum Stress Criterion,* but originally denoted as "quantum macro-mechanics of fracture" (Petrov, 1991,1996)):

$$\sigma_d^* = \langle \langle \sigma_{tip} \rangle_0^{\Delta l} \rangle_{t-\Delta t}^t = \sigma_C \quad \text{Mode I; for Modes II, III: } \sigma \rightarrow \tau \qquad (4)$$

Such stress-based criteria, originally proposed for mode I, can be obviously simple extended for modes II and III as reported in eqs. (3) and (4).

Note that imposing that the criteria of eqs. (1) and (3) predict the same failure stress corresponds to a coupled stress and energy failure criterion (Cornetti, 2004; Cornetti, Carpinteri, Pugno and Taylor, 2004; a similar but different treatment was proposed by Leguillon, 2002), i.e.:

$$K^* = K_C \quad \& \quad \sigma^* = \sigma_C \quad : \text{ same predictions (or } \sigma \rightarrow \tau \text{ for Modes II,III)} \qquad (5)$$

Correspondingly, imposing the same predictions from eqs. (2) and (4) a new dynamic *Quantized Coupled Criterion for unstable crack propagation* (in which the fracture and time quanta are derived to ensure the equality of such predictions) is formulated:

$$K_d^* = K_C \quad \& \quad \sigma_d^* = \sigma_C \quad : \text{ same predictions (or } \sigma \rightarrow \tau \text{ for Modes II,III)} \qquad (6)$$

In addition, substituting the corresponding strain in eqs. (3) and (4), normal ($\varepsilon$) for mode I or tangential ($\gamma$) for mode II and III, a *Quantized Maximum Strain Criterion* (Pugno and Ruoff, 2004a):

$$\varepsilon^* = \left\langle \varepsilon_{tip} \right\rangle_0^{\Delta l} = \varepsilon_C \quad \text{Mode I;} \quad \text{for Modes II, III: } \varepsilon \to \gamma \tag{7}$$

and a new *Dynamic Quantized Maximum Strain Criterion*:

$$\varepsilon_d^* = \left\langle \left\langle \varepsilon_{tip} \right\rangle_0^{\Delta l} \right\rangle_{t-\Delta t}^{t} = \varepsilon_C \quad \text{Mode I;} \quad \text{for Modes II, III: } \varepsilon \to \gamma \tag{8}$$

are derived.

The criteria of eqs. (3-8) require the expression of the complete (and not only asymptotic) stress field around the tip of the defect (to be more powerful than the corresponding classical ones), well-known only for the simplest cases. On the other hand, the criteria of eqs. (1,2) can be applied in a very simple way by starting from the well-known solutions for the stress-intensity factors (e.g., for the quasi-static case see the Murakami's Handbook, 1986). Obviously, the predictions of the different criteria are different but similar (a comparison between eqs. (1), (3) and (7) for predicting the strength of defective nanotubes was reported by Pugno and Ruoff (2004a)).

For the Griffith's case (linear elastic infinite cracked plate under remote tension $\sigma$ orthogonal to the crack -of length $2l$) $K = \sigma\sqrt{\pi l}$ (mode I) and thus $K^* = \sigma\sqrt{\pi(l + \Delta l/2)}$; $K^* \approx K$ only for very large cracks. Denoting with $\Delta K_C$ the threshold value of the stress-intensity factor in fatigue, for very large cracks $\Delta K_C \approx \Delta K_C^*$, whereas for very short cracks $\Delta\sigma_C \approx \Delta K_C / \sqrt{\pi \Delta l/2}$), where $\Delta\sigma_C$ is the plain-specimen fatigue limit. This yields an estimation of the fracture quantum during fatigue crack propagation as $\Delta l \approx \dfrac{2\Delta K_C^2}{\pi \Delta\sigma_C^2}$. Thus in general the fatigue limit $\Delta\sigma_f$ for a cracked large plate is predicted as $\Delta\sigma_f \approx \Delta K_C / \sqrt{\pi(l + \Delta l/2)}$, exactly as experimentally observed (see Taylor, 1999 and Taylor, Cornetti and Pugno, 2004).

Accordingly, eqs. (1) and (3) were rewritten (Taylor, 1999 and Taylor, Cornetti and Pugno, 2004, respectively) for fatigue limit predictions, formally considering the variations $\Delta$ before the stresses and the stress-intensity factors. On the other hand, here we note that all the eqs. (1-8) can be rewritten in the same manner, formulating new criteria for fatigue limits (also with rapid alternating loading): *static and dynamic* (i) *maximum variation of the stress-intensity factor,* (ii) *of the stress,* (iii) *of the strain and* (iv) *coupled criteria for fatigue limits*.

On the other hand, regarding the evolution of the fatigue crack, substituting the stress-intensity factor $K$ with its "quantized" version $K^*$ in the Paris' law, we formulate a new quantized fatigue crack growth law, to be applied also to short cracks (*Quantized Paris' Law*):

$$\frac{dl}{dN} \approx C\left(\Delta K^*\right)^m \tag{9}$$

where $N$ is the number of cycles, C, m are the Paris' constants and $\Delta K^*$ is the variation of the "quantized" stress-intensity factor in a cycle. This yields in addition an elegant interpretation of the threshold value for $\Delta K$, as connected to the existence of the fracture quantum, i.e., $\Delta K_C^* \approx \Delta l^{1/m}/C$ ($\Delta l$ during $\Delta N = 1$).

For very short cracks $\Delta K^* \propto \Delta\sigma$ and eq. (9) resembles the classical Whöler's law, i.e., $N_f(\Delta\sigma)^{\overline{m}} = \overline{C}$ ($\overline{C}, \overline{m}$ constants; $N_f$ life time), as well as for very large cracks $\Delta K^* \approx \Delta K$ and eq. (9) becomes the classical Paris' law. Eq. (9) corresponds for fatigue (stable crack propagation) to eq. (1) for brittle crack propagation. The analog of eq. (3) can be formulated for fatigue substituting $\Delta\sigma$ with $\Delta\sigma^*$ in the classical Whöler's law. Accordingly, starting from these two analogs in fatigue, it is clear that all the analogs of eqs. (1-8) can be easily formulated also for fatigue crack growth.

## 3. Do "materials become insensitive to flaws at nanoscale"? In general not. The example of carbon nanotubes

By applying eq. (1) to an infinite plate with a "predominant" symmetric crack of half-length $l$ and blunt tip radius $\rho$, we derive the strength $\sigma_f$ of the plate as (Pugno and Ruoff, 2004a):

$$\sigma_f = K_{IC}\sqrt{\frac{1+\rho/(2\Delta l)}{\pi(l+\Delta l/2)}} = \sigma_C\sqrt{\frac{1+\rho/(2\Delta l)}{1+2l/\Delta l}} \qquad (10)$$

where $K_{IC}$ is the fracture toughness and $\sigma_f(l=\rho=0) = \sigma_C$ is by definition the strength of the material for the plain structure; in general, it differs from the ideal strength since other minor defects could exist in the plate.

Particularizing eq. (10) to the case of sharp cracks, i.e., $\rho = 0$, the same strength prediction is obtained by applying eq. (3) and thus also by applying eq. (5), whereas (i) classical fracture mechanics would yield eq. (10) with $\Delta l \to 0$ (and $\rho = 0$) predicting an infinite strength for the plain structure, i.e., $\sigma_f(l=0) = \infty$, clearly a paradox; on the other hand, (ii) the classical maximum stress criterion simply would imply a vanishing strength, i.e., again a paradox. In contrast, eq. (10) does not present paradoxes. It unifies stress-concentration and -intensification factors. It suggests that for $2l << \Delta l$ (very short cracks) materials become insensitive to flaws, as observed by Gao et al., (2003) in nano-biocomposites: for this case the fundamental critical parameter is $\sigma_C$, whereas for very large cracks it becomes $K_{IC}$. Making an analogy, this explains why similar phenomena, but arising at different size scales, as fracture and wear, are governed by different competing parameters (respectively $K_{IC}$ and hardness $H \propto \sigma_C$).

Gao et al. (2003) predict an insensitive to flaws not for small crack lengths but for small structural sizes. However, we note that this difference is only formal and not substantial since small structures cannot contain large cracks. In addition, we note that their approach was previously introduced by Carpinteri (1982) in the context of the competition between brittle and tensional collapses, even if not specifically in the field of nanomechanics (for which the ultimate strength has simply to be replaced by the ideal material strength).

More importantly, the fracture quantum is itself a size dependent parameter (that increases by increasing the size scale) as suggested by its prediction in brittle fracture, i.e., $\Delta l \approx \frac{2 K_{IC}^2}{\pi \sigma_C^2}$, see eq. (10), in which we must remember that $\sigma_C$ is the strength of the plain structure (that increases by decreasing the size scale as a consequence of approaching the defect-free condition, for which $\sigma_C$ becomes coincident with the ideal material strength). At macroscale the fracture quanta for brittle and fatigue crack propagation are different (since usually $\frac{K_C}{\Delta K_C} > \frac{\sigma_C}{\Delta \sigma_C}$, see Ciavarella 2002), but at nanoscale they could become coincident (conjecture) to the distance between adjacent chemical bonds (as well as the time quantum is expected to be the finite time to generate a fracture quantum). For example, for nanotubes the fracture quantum for brittle fracture truly becomes identical to the distance between adjacent chemical bonds, as demonstrated by comparing eq. (10) with atomistic simulations of various types (see Pugno and Ruoff, 2004a). Thus, for that nantoubes, even the smallest defect, e.g., just a simple vacancy, affects considerably the strength. For example, eq. (1) was applied to demonstrate that just one vacancy reduces the strength of a nanotube (or of a two dimensional atomic lattice) by a factor of ~20%. Molecular mechanics (MM) and quantum mechanical calculations agree with this prediction. Thus, assuming defects as adjacent vacancies, the band between ~80% and ~100% of the ideal strength is "forbidden": the strength is quantized as a consequence of the quantization of the defect size.

For example, let us consider for nanotubes the fracture quantum identical to the distance between two adjacent broken chemical bonds, i.e., $\Delta l \approx \sqrt{3} a$, with $a \approx 1.42 \text{Å}$ interatomic spacing. Considering defects like $n$ adjacent vacancies, $2l = n \Delta l$ in eq. (10). This case was also treated by MM atomistic simulations for an (80,0) carbon nanotube. The MM-calculated strengths clearly follow the $(1+n)^{-1/2}$ quantization predicted by eq. (10) with a fit of $\sigma_C \sqrt{1 + \rho/2\Delta l} = 111 \text{GPa}$. From the value of the ideal strength calculated by MM, $\sigma_C = 93.5 \text{GPa}$, the reasonable value of $\rho \approx 0.8 \Delta l \approx 2.0 \text{Å}$ was deduced. For $n$=2 QFM (eq. (10)) predicts 64.1GPa against 64.1GPa predicted by MM; for $n$=4, 49.6GPa (QFM) against 50.3GPa (MM); for $n$=6, 42.0GPa (QFM) against 42.1GPa (MM); for $n$=8, 37.0 (QFM) against 36.9GPa (MM). Note that experiments on fracture strength of carbon nanotubes (Yu et al., 2000) emphasized clusters at 63GPa (close to the prediction for $n$=2), 43GPa (close to the prediction for $n$=6) and 39GPa (close to the prediction for $n$=8); for details please refer to Pugno and Ruoff (2004a).

These examples clearly show that "materials become insensitive to flaws at nanoscale", as reported in the title of the interesting paper by Gao et al. (2003), is in general not true. As previously emphasized the reason is that a "crack insensitiveness zone" exists but only for flaws smaller than the fracture quantum $\Delta l$, and for nanotubes we have demonstrated that $\Delta l = \sqrt{3} a \approx 2.5 \text{Å}$ ! Obviously, a vacancy in a large real (thus defective) object does not affect its strength, since at such a scale the fracture quantum will be larger, as suggested by its expression, in which $\sigma_C$ is expected to be much smaller than the ideal strength of the material. In other words, other defects will predominate.

## 4. Increasing of the dynamic strength

Let us consider as a simple example a semi-infinite crack in an otherwise unbounded body. The body is initially stress free and at rest. At time $t=0$ a pressure $\sigma$ begins to act on the crack faces. In this case, as it is well known, $K_I(t) = 2\sigma \dfrac{\sqrt{c_D t(1-2\nu)/\pi}}{(1-\nu)}$ (see Freund, 1990), where $c_D$ is the dilatational wave speed of the material and $\nu$ is its Poisson's ratio. Applying eq. (2) we find the failure for a given time $t_f > \Delta t$, satisfying:

$$2\sigma \frac{\sqrt{c_D(1-2\nu)/\pi}}{(1-\nu)} = \frac{K_{IC}}{\sqrt{t_f - \Delta t/2}} = \frac{1}{\sqrt{t_f}} \frac{K_{IC}}{\sqrt{1 - \Delta t/(2t_f)}} = \frac{K_{dIC}}{\sqrt{t_f}} \qquad (11)$$

Note that, according to our time quantization, a minimum time to failure exists and it must be of the order of $t_{f\min} \approx \Delta t$. In addition, if classical dynamic fracture mechanics is applied ($\Delta t = 0$), the "measured" fracture initiation toughness $K_{dIC}$ will be observed, according to dynamic quantized fracture mechanics (eq. (11)), time to failure dependent. In fact, considering very severe impacts ($t_f \to t_{f\min} \approx \Delta t$), the dynamic strength ($\propto K_{dIC}$) is expected for this scheme $\sqrt{2}$ times larger than its static value ($\propto K_{IC}$). Considering an applied pressure linearly increasing with time, the factor $\sqrt{2}$ is replaced by the factor 2 (Pugno and Ruoff, 2004b). This increasing of the dynamic strength has been observed experimentally (Owen at al., 1998) on microsecond range dynamic failures of 2024-T3 aircraft aluminum alloy, where the dynamic strength was observed increasing by a factor of ~2 by varying the time to failure by ~8 order of magnitudes. Also, Owen at al. (1998) reported the observation of a minimum time to failure (see Pugno and Ruoff, 2004b for details and for the comparison with the Petrov's criterion of eq. (4)).

## 5. Micro-experiments on tensile strength: the example of the UNCD

In order to investigate the strength of freestanding UNCD thin films at microscale, the membrane deflection experiment was considered (Espinosa et al., 2003; Pugno, Peng and Espinosa 2004; Peng, Espinosa and Moldovan, 2004; Peng, Pugno and Espinosa, 2004). The technique involves the stretching of freestanding specimens with micron thickness in a fixed-fixed configuration. The specimen geometry utilized by such a technique resembles the typical dog-bone tensile specimen, but with an area of additional width in the specimen center, designed to prevent failure at the point of application of a line load. The suspended membranes are fixed to the wafer at both ends such that they span a bottom view window. In the areas where the membrane is attached to the wafer and in the central area the width is varied in such a fashion to minimize boundary-bending effects. These effects are also minimized through large specimen gauge lengths. Thus, a line load applied by a nanoindenter in the center of the span results in direct stretching under large displacements of the membrane (as would be for a cable) in the two areas of constant width as in a direct tension test. Simultaneously, an interferometer focused on the bottom side of the membrane records the deflection. The result is direct tension of the gauged regions, in the absence of strain gradients, with load and deflection being measured independently. The data directly obtained from the experiment is processed to arrive at a stress-strain signature for the membrane. In addition, the interferometer yields vertical displacement information in the form of monochromatic images taken at periodic intervals. The relationship for the distance between fringes, is related through the wavelength $\lambda$ of the monochromatic light used. Assuming that the membrane is deforming uniformly along its gauge length, the relative deflection between two points can be calculated, independently from the nanoindenter measurements, by counting the total number of fringes and multiplying by $\lambda/2$ (Espinosa et al., 2003; Pugno, Peng and Espinosa 2004).

An important aspect of the UNCD specimens was that each membrane bowed upward as processed, i.e., out of the wafer plane. This is believed to result from the difference in thermal expansion coefficients between the film and Si wafer such that cooling down from the deposition temperature, approximately 800°C, resulted in the Si shrinking more than the UNCD film. The film curvature is indicative of a gradient of residual stresses across the film thickness. The out-of plane profile was obtained through the interferometric measurements (Espinosa et al., 2003). From this profile, the height above the plane of the wafer was determined. Also, the profile was used to measure the actual length of the curved membrane, which is used to determine the downward deflection, corresponding to the beginning of uniform specimen straining, after the snap-through instability. This rather general experimental procedure can be applied to different materials (Peng, Pugno and Espinosa, 2004). It emphasized significant size-effects on UNCD strength, that will be discussed in the following section.

## 6. Size effects at microscale as an indirect observation for the nanoscale strength: the example of the UNCD

Peng, Espinosa and Moldovan (2004) have measured the strength of UNCD membranes of 1μm thick with width/length of 5/100, 10/200, 20/200 and 40/400 microns, also with nitrogen gas added of 5,10,15 and 20% in the atmosphere. Thirty tests were performed for specimens with specified doping and size, for a total of 480 tests, making it possible to apply statistical concepts. In particular, we apply fractal statistics (Carpinteri and Pugno, 2002, 2004a) to these results for predicting the still unmeasured strength of UNCD at nanoscale. The coefficient of correlations are found basically identical to 1, showing that fractal statistics can be applied with confidence to these tests.

Failure can be also described by the widely used Weibull's (1939) statistics. Weibull statistics allow examination of strength (or time to failure or fatigue life) in the sense of failure probability at a certain stress level. The simplest Weibull distribution is defined as $P_f = 1 - \exp\left(-(R/R_0)^D (\sigma_f/\sigma_0)^m\right)$, where $\sigma_f$ is the failure stress, $\sigma_0$ is the stress scaling parameter: in other words, it is the nominal stress that would result in 63% (i.e., $(1-e^{-1}) \cdot 100\%$) of the specimen to fail, having characteristic size $R$, i.e., volume $V = R^3$; $D = 3$ if we classically assume volume predominant defects. For predominant surface defects $A = R^2$ and thus $D = 2$ ($A$ is the specimen surface area); $m$ is the Weibull modulus, which can be identified from a log-log plot of the probability of failure; $R_0$ is the reference size on which the Weibull parameters are identified. It is not clear if volume or surface has to be considered, even if at small scale surface defects should become predominant. Accordingly, the size-effect on the strength $\sigma_f$ (at a specified $P_f$) is predicted as $\sigma_f \propto R^{-D/m}$.

We finally note that Weibull statistics has to be rewritten in its integral form, i.e., $P_f = 1 - \exp\left(-\frac{1}{V_0} \int_V (\sigma/\sigma_0)^m dV\right)$, if stress gradients are present in the structure (e.g., for bending). Moreover, if stress-intensifications are present (e.g., cracked structures) the previous integral does not converge: this represents a limit of the classical Weibull statistics and can be automatically removed if instead of $\sigma$ its "quatized" version $\sigma^*$ (or $\sigma_d^*$) is considered: the quantized crack advancement removes an other paradox.

Based on the fractal statistics (Carpinteri and Pugno, 2002, 2004a), that assumes energy dissipation in a fractal domain of dimension $D - 1 \leq \mathcal{D} \leq D$, e.g., comprised between Euclidean surface ($D$=2) and volume ($D$=3) if a three-dimensional object ($D$=3) is considered (here the size-effect on Young's modulus is neglected), we have:

$$\sigma_f \propto R^{\frac{\mathcal{D}-D}{2}}, \quad D - 1 \leq \mathcal{D} \leq D, \quad D = 2,3 \tag{12}$$

Thus, we can demonstrate the equivalence between the size effects predicted by the Weibull and fractal statistics (if the Young's modulus scaling is neglected) in terms of:

$$m = \frac{2D}{D - \mathcal{D}} \tag{13}$$

Note that eq. (12) can be applied also to one-dimensional objects, for which $D = 1$.

Regarding size-effects it is interesting to note that integrating eq. (9) (with m>2) just simply assuming the Griffith's case and initial crack length proportional to the structural dimension $R$ and much larger than the fracture quantum, would correspond to $\sigma_f \propto R^{\frac{1}{m}-\frac{1}{2}}$, giving consequently the correlation (based on the previous hypotheses) $m \approx \frac{2}{D - \mathcal{D} + 1}$ between Paris' and fractal exponents. For the classical case of fractal exponent equal to the corresponding Euclidean dimension, i.e., $\mathcal{D} = D$, corresponding to vanishing size-effects, we find $m = 2$, in agreement with the classical models developed for interpreting the Paris' law (see Ciavarella, 2002), that in fact do not consider the size effect on material strength.

The advantage of the fractal statistics, eq. (12), is the clear interpretation of $D$, whereas the physical meaning of $m$ remains partially unclear. If the fractal approach is correct, the experimental size effect must give good fits with fractal exponents in the ranges $(D-1, D)$. Assuming the investigated microspecimens as two-dimensional objects, i.e., as thin films (energy dissipations invariable along the thickness $t$, $D = 2; R = \sqrt{WL}$), the observed size-effects on doped and undoped UNCD microfilms compared with the fractal statistics are depicted in Figure 1a showing a very good fit. If the specimen is considered as a three-dimensional microbeam (energy dissipations can vary along the thickness $t$, $D = 3; R = \sqrt[3]{WLt}$) the interpretation is different (but here similar as a consequence of the constant thickness of the specimens), as shown by the comparison reported in Figure 1b; in both the cases the fractal approach seems to be consistent, showing a very good fit (the coefficient of correlation is basically identical to 1 for all the fits) with $D-1 \leq D \leq D$. Clearly the doping decreases the fractal dimension of the energy dissipation. Even if the UNCD electrical conductivity is strongly increased by the presence of nitrogen, its strength at microscale is strongly reduced by the presence of the doping (Peng, Espinosa and Moldovan, 2004). Something different could happen at nanoscale.

On the basis of the fractal statistics an estimation of the strength of UNCD at nanoscale can be derived. For example, considering a UNCD nanowire ($D = 3; R = \sqrt[3]{WLt}$) with $W=t=$10nm, $L=$100nm, the fractal statistics predicts a strength for the undoped UNCD of ~23GPa (against the maximum value observed at microscale of ~5GPa), with fractal exponent $D=2.48$ (and Weibull modulus $m=11.6$, from eq. (13)); for 5% doping the nanoscale strength is ~22GPa and $D=2.37$ ($m=9.5$); for 10% the strength is of ~28GPa and $D=2.25$ ($m=8.0$); finally for 20% doped UNCD the strength is of ~34GPa (against the maximum value observed at microscale of ~3GPa) and $D=2.20$ ($m=7.5$).

For cantilever nanowires the nanoscale bending strength $\sigma_{fB}$ is also of interest. It is expected to be larger than the corresponding tensile strength $\sigma_f$ as a consequence of the reduced volume undergoing larger stresses. According to Weibull (1939) $\sigma_{fB} = \sigma_f (2m+2)^{1/m}$. Thus, we expect nanoscale bending strengths of about 30, 30, 40 and 50 GPa respectively for 0, 5, 10 and 20% UNCD doping.

We note that the strength for doped nanowires is predicted larger than for the undoped nanowire. This last consideration has to be taken with caution (see concluding remarks); however, it would be a consequence of the lower fractal dimension imposed by the presence of the nitrogen, resulting in a larger negative slope for the doped curves in Figures 1 with respect to the undoped ones.

## 7. Secondary crack emanation and fractal crack surface formation

If a fractal nature for the energy dissipation during a crack advancement $\Delta l' > \Delta l$ is assumed, for example due to the formation of a fractal crack surface (Carpinteri and Chiaia, 1996) and/or due to the emanation of secondary fractal cracks, the "apparent" fracture energy $G_C'$ (dissipated per unit area created) will refer to the increasing $\Delta l'$ of the nominal crack length. The energy equivalence $G_C \Delta L = G_C' \Delta l'$ must hold, where $G_C$ and $\Delta L$ are the real fracture energy and the real total crack length increment. The smallest crack length coincides with the fracture quantum $\Delta l$, so that $\Delta L = (\Delta l'/\Delta l)^{D-1} \Delta l$ (Kashtanov and Petrov, 2004), where $D$ is the fractal exponent describing the fractal nature of the crack. Accordingly, we derive:

$$G_C' = G_C \left(\frac{\Delta l'}{\Delta l}\right)^{D-\mathrm{D}+1} \tag{14}$$

where usually D, the topologic dimension of the object, is equal to 3 (or 2) and $D$ belongs to the range (2,3) (or (1,2)). In this case eq. (14) predicts $G_C' = G_C$ only if the classical Euclidean crack surface, i.e., $D$=2, is considered, whereas for larger value of $D$, describing a fractal surface area, $G_C' > G_C$.

## 8. Estimation of the ideal strength by Quantized Fracture Mechanics: the example of the UNCD

The computed nanoscale UNCD strength is compatible with a simple estimation of the UNCD ideal strength $\sigma_C$, obtained by applying eq. (1). We expect an even higher value for the UNCD ideal strength, as suggested by the observed size effects (smaller is stronger). Assuming the fracture quantum coincident with the grain size $d$ (~3nm) of the UNCD, the ideal strength is estimated according to eq. (1) as:

$$\sigma_C \approx \frac{K_{IC}}{\alpha}\sqrt{\frac{2}{\pi d}} \tag{15}$$

where $K_{IC} \approx 4\mathrm{MPa}\sqrt{\mathrm{m}}$ is the fracture toughness of the UNCD and $\alpha$ is a parameter equal to 1 for structures "without free surface" (e.g., the previously treated infinite plate) or equal to 1.12 for structures "with free surface" (e.g., finite plate), that takes into account the "edge effect". According to eq. (15) the surfaces are predicted weaker than the inner parts of a solid by a factor of ~10%. For finite structures, we find a reasonable estimation of the ideal UNCD strength of ~52GPa. If the structure is assumed infinite the result would be ~58GPa. The size $R$ in eq. (12) corresponding to a strength equal to the ideal strength of the solid fixes the limit size that can be treated by the fractal approach. From eq. (15) $\sigma_C \propto d^{-1/2}$ (in agreement with the well-known Hall-Petch relationship), that suggests nanostructured materials (small grain size $d$) for high strength applications; however, we have to note that the constant of proportionality (not specified by the Hall-Petch relationship) is basically $K_{IC}$ and could decrease itself by decreasing the grain size (here we assume $K_{IC}$ as a constant), limiting the fracture toughness and the applicability of nanostructured materials. Thus, the fracture toughness more than the strength could be the real critical point in designing nanostructured materials. Consequently, zones with high stress-concentrations and -intensifications, as for example surface steps (e.g., re-entrant corners, see Pugno, Peng and Espinosa, 2004; Carpinteri and Pugno, 2004b), must be avoided (e.g., with high quality surface polishing).

## 9. Concluding remarks

Summarizing the quantized criteria proposed in Section 2 and especially that ones of eqs. (1,2,9) based on stress-intensity factors, reported in Handbooks for hundred of cases, are useful tools for the predictions of the strength of defective solids, in both static, dynamic, stable (brittle) and unstable (fatigue) crack propagations, also at nanoscale. In Section 3 we have demonstrated that solids can be sensitive to flaws (also) at nanoscale, see eq. (10) in which for example for nanotubes the fracture quantum $\Delta l$ has been demonstrated to become of the order of the atomic spacing. In Section 4 the increasing of the dynamic strength has also been rationalized.

From the fractal statistics, eq. (12), the nanoscale strength of doped and undoped UNCD is estimated; thus, it is demonstrated that the proposed methodology can be applied in general for estimating the nanoscale (tensile and bending) strength of materials, experimentally investigated at the more accessible microscale. Weibull statistics would give the same prediction, on the basis of the correlation of eq. (13).

The role of a fractal secondary crack emanation and/or crack surface formation is also described, according to eq. (14).

A simple application of eq. (1) is demonstrated to be able to give estimations for the ideal material strength, eq. (15). Accordingly, the surfaces are predicted weaker than the inner parts of a solid by a factor of ~10%. Obviously the extrapolation of the nanoscale strength from eq. (12) is based on the assumption that the dimension of the fractal domain (exponent in eq. (12) or slope of the straight lines in Figures 1) can be considered as a constant from micro- to nano-scales. And this could be not fully verified especially for the more complex case of doped UNCD specimens. Similarly, the validity of the simple approach summarized in eq. (12) (that assumes fracture quantum identical to the grain size for UNCD, assumed in addition identical to 3nm) is approximated. Thus, the UNCD predictions here reported must be considered simple reasonable estimations. Detailed quantum mechanical calculations are needed for deriving better predictions.

However, the described quantized criteria and statistical/experimental procedure are in general useful tools in the study of the strength of solids, also at nanoscale.

**Acknowledgement**

The author would like to acknowledge Profs. A. Carpinteri, R. Ruoff, Y. Petrov, I.V. Simonov, P. Cornetti, N. Moldovan and H. Espinosa for the discussion. And especially Diane Dijak for the English grammar supervision.

**FIGURE CAPTIONS**

Figure 1:

(a) Comparison between fractal statistics (straight lines) and experimental size-effects (data points) on UNCD strength, for undoped (fractal exponent $D$=1.97) and nitrogen doped (5%, $D$=1.58; 10%, $D$=1.50; 20%, $D$=1.47) specimens at microscale; such specimens are here treated as two-dimensional structures, i.e., as thin films.

(b) Comparison between fractal statistics (straight lines) and experimental size-effects (data points) on UNCD strength for undoped (fractal exponent $D$=2.48) and nitrogen doped (5%, $D$=2.37; 10%, $D$=2.25; 20%, $D$=2.20) specimens at microscale; such specimens are here treated as three-dimensional structures, i.e., as microbeams.

**FIGURES**

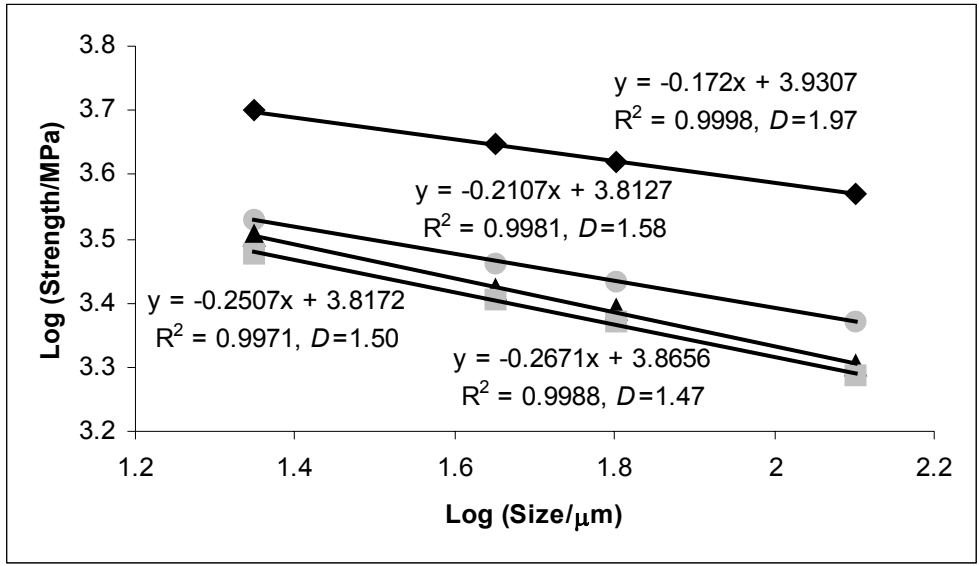

Figure 1a

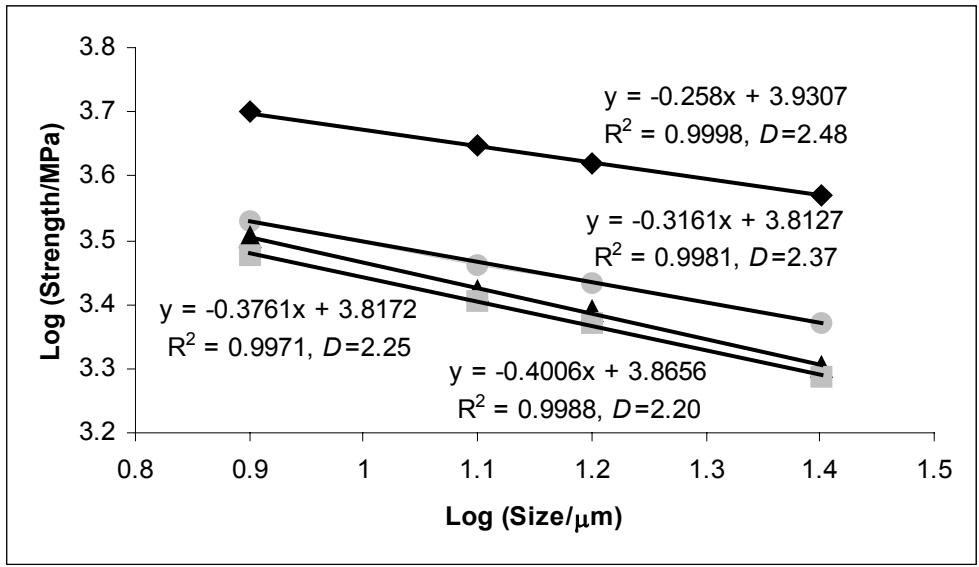

Figure 1b